\documentclass[12pt]{article}
\usepackage{cite}
\usepackage{pifont}
\usepackage{times} 
\usepackage{amssymb,amsbsy}
\usepackage{amsmath}
\usepackage{graphics}
\usepackage{pstricks}
\newcommand\email[4]{#1@#2.#3.#4}
\newcommand\tr{\operatorname{Tr}\,}
\def\C{{\mathbf{C}}}

\def\Z{{\mathbf{Z}}}
\def\w{{\lambda}}

\def\eq#1{(\ref{#1})}

\def\viz{\emph{viz.}}
\newcommand\pa{\partial}

\newcommand\thetfn[4]{\vartheta\genfrac{[}{]}{0pt}{2}{{\scriptstyle #1}}{{\scriptstyle #2}}({#3},{#4})}
\title{Partition function of beta-gamma system on orbifolds}
\author{
Chandrasekhar Bhamidipati\thanks{\email{chandrasekhar}{iitbbs}{ac}{in}} \\
\small School of Basic Sciences, 
\small Indian Institute of Technology Bhubaneswar \\
\small Bhubaneswar 751~007, India.
\and 
Koushik Ray \thanks{\email{koushik}{iacs}{res}{in}} \\
\small Department of Theoretical Physics,
\small  Indian Association for the Cultivation of Science \\
\small  Calcutta 700~032. India.
}
\date{}
\begin{document}
\setcounter{page}{1}
\maketitle
\thispagestyle{empty}
\begin{abstract}
\noindent 
Partition function of beta-gamma systems on the orbifolds
$\C^2/\Z_N$ and $\C^3/\Z_M\times\Z_N$ are obtained
as the invariant part of that on the respective affine spaces,
by lifting the geometric action of the orbifold group
to the fields. Interpreting the sum over roots of unity as 
an elementary contour integration, the partition function 
evaluates to an infinite series counting invariant monomials composed of
basic operators of the theory at each mass level.
\end{abstract}
\section{Introduction}
A curved beta-gamma ($\beta$-$\gamma$) system is defined as the chiral sector 
of a certain infinite radius limit of the two-dimensional non-linear sigma 
model on a complex variety as the target space \cite{wit1,nek1}.  
Since its incipience in the study of
sheaves of vertex operators of chiral de Rham complexes
\cite{Malikov:1998dw,Malikov:1999ab},
beta-gamma systems have been studied in the context of topological strings 
\cite{kap,Berkovits:2005bt}, geometry of $(0,2)$ sigma-models 
\cite{wit1,Tan:2006qt,Tan:2006by,aisaka,gp,Grassi:2007va}, 
mirror symmetry \cite{frenkel} and pure spinors 
\cite{berkovits,arroyo,morales,aabn}. 
In the pure spinor formalism, for instance, 
the non-linear beta-gamma theory appears as a ghost system whose
partition function is  evaluated as characters
of pure spinors \cite{Berkovits:2005hy}. 
The moments of the partition function of the zero
modes is identified with the multiplicities of the ghosts which 
in turn provide the Virasoro central charge and are related to the current 
algebra of ghosts. 

In this article we consider a beta-gamma system on orbifolds of
two and three-dimensional complex affine spaces, namely $\C^2/\Z_N$ and
$\C^3/\Z_M\times\Z_N$. 
The partition function on an affine space is obtained as the generating
function of chiral operators, namely the fields $\gamma$, identified as 
the complex coordinates of the target space, their conjugates and their
world-sheet derivatives, graded by the scaling degree of the fields
and mass. Defining a beta-gamma system on a curved space requires identifying 
the fields $\gamma$ with the coordinates of the curved space, which can only be
effected on each affine chart. Coordinate transformations made compatible
with the operator products provide the rules for changing patches \cite{nek1}.
If a variety is given as a set of equations in the coordinate ring of an affine 
or projective space, defining  a beta-gamma system entails imposing the 
equations as constraints. This is the case for pure spinors, for example. 
The orbifolds we consider may also be defined in this fashion. Equivalently,
however, the geometric action of the orbifold group  
can be lifted to an action on the fields $\gamma$
through its identification  with the coordinates of the affine
space. The scaling symmetry then determines the action on the conjugates 
$\beta$. We exploit this for restricting the generating function of chiral
operators to the sector invariant under the orbifold group to obtain 
the partition function on the orbifolds.

Restricting the generating function to the invariant sector entails a sum
over roots of unity. This is achieved using rudimentary contour integration 
in a single complex variable and leads to a series in powers of
the modular parameter. The massless or zero-mode part of the partition 
function can be easily extracted from the series.
It arises from the invariant monomials of the orbifold group in 
the coordinate ring of the affine variety. The generating function is then
its Molien series \cite{stanley,smith}. For simple cases, the Molien series
can also be obtained from the
syzygies defining the orbifold in the coordinate ring of the ambient affine
space. We show that the zero mode part of the
partition function of the beta-gamma system matches with the Molien series.
In principle, evaluating all the moments of the Molien series enables one to
write the partition function of the non-zero modes as well. However,
evaluation of infinite number of moments presents practical difficulties in
implementing this procedure. A series is better suited for explicit 
evaluation of the partition function.
\subsection{Partition function of beta-gamma system on $\C^d$}
Let us start with a discussion of the partition function of 
a beta-gamma system on a complex affine space.
A beta-gamma system refers to a two-dimensional conformal field theory
with a set of complex fields $\{\gamma^i\}$ and their canonical conjugates, 
$\{\beta_i\}$, $i=1,2,\cdots , d$. The fields $\gamma$ have vanishing
conformal dimension and  $\beta$ are
one-forms on the two-dimensional space-time, referred to as the  
world-sheet, namely, $\beta_i = \beta_{i\xi}
d\xi+\beta_{i\bar{\xi}}d\bar{\xi}$, $\xi$ denoting the complex 
coordinate of the world-sheet and a bar the complex
conjugate. Since the complex affine space $\C^d$ 
can be covered by a single coordinate chart, 
the fields are identified with the coordinates
as $\gamma^i=x_i$ of the coordinate ring of $\C^d$, \viz
$\C[x_1,x_2,\cdots , x_d]$ and are assigned the free operator product
\begin{equation}
\gamma^i(\xi) \beta_j(\xi') \sim \delta^i_j\frac{d\xi'}{\xi-\xi'}
\end{equation} 
with their conjugates. 
The theory is described by an action written in the conformal gauge as 
\begin{equation}
S = \frac{1}{2\pi}\int \beta_i\bar{\pa}\gamma^i, 
\end{equation} 
where $\pa = \frac{\pa}{\pa\xi}$. The corresponding equations of motion
restrict the fields to be holomorphic on the world-sheet. 
In considering an orbifold of the affine space,
the identification of $\gamma$'s with the coordinates
allows lifting the action of orbifold group to the fields. 

The theory possesses two conserved currents, the energy momentum tensor
 and an $U(1)$ current arising from scaling of the fields as
\begin{equation}
\label{eq:scaling}
\gamma^i\longrightarrow \Lambda_i\gamma^i, \quad
\beta_i\longrightarrow\Lambda_i^{-1}\beta_i.
\end{equation} 
The parameters $\Lambda_i$ acting on the different fields are not
independent. All of them are functions of a single parameter, ergo the
symmetry group is not $U(1)^d$ but $U(1)$.
The respective charges, namely, 
 $L_0=\oint d\xi \xi \beta_{i\xi}\partial\gamma^i$ and
$J_0 = \oint d\xi \beta_i\gamma^i$, 
characterize the field theory.  Introducing the modular parameter 
$q$ and another
parameter $t$ corresponding to the scaling, then, the partition function of
the beta-gamma system is written as 
\begin{equation}
\mathcal{Z} = \tr (q^{L_0} t^{J_0}), 
\end{equation} 
where $\tr$ signifies a trace with respect to the states of the 
Hilbert space of the theory. 

With the identification alluded to above,
the partition function of the beta-gamma system on $\C^d$ evaluates to
\cite{gp}
\begin{equation}
\label{zcd}
\mathcal{Z}_{\C^d} = (\mathcal{Z}_{\C})^d, 
\end{equation} 
where
\begin{equation}
\label{zc}
\mathcal{Z}_{\C} = \frac{1}{1-t}\prod_{n=1}^{\infty} 
\frac{1}{(1-q^nt)(1-q^n/t)}.
\end{equation} 
Let us interpret this combinatorially. 
First, let us consider the one-dimensional
case $\C$, with coordinate ring $\C[x]$.
The partition function is the generating function of products of the fields
$\gamma$, $\beta$ and their derivatives graded by their 
$q$-degree and $t$-degree, given by the charges $L_0$ and $J_0$,
respectively. Each power of $x=\gamma$ furnishes a $t$
and each world-sheet
derivative $\pa$ a $q$. The scaling \eq{eq:scaling} dictates
each $\beta$ to contribute $1/t$ while being an one-form it also furnishes 
a factor of $q$, that is, $q/t$, in total. These contributions are collected 
in Table~\ref{tab:deg}.
\begin{table}[h]
\centering
\begin{tabular}{rc}
\hline\hline
object & degree\\
\hline
$x$ or $\gamma$  & $t$\\
$\pa$ & $q$\\
$\beta$ & $q/t$\\
\hline
\end{tabular}
\caption{$q$- and $t$-degrees of fields}
\label{tab:deg}
\end{table}
A monomial constructed with different powers and combinations 
of $x$ or $\gamma$, $\pa$ and  $\beta$ furnish an
object with a certain $q$- and $t$-degrees. For example, objects of vanishing
$q$-degree, namely, the zero modes, arise only 
from integer powers of $x$; the $t$-degree of a monomial
$x^n$ for any positive integer $n$ is $n$. 
As for objects of higher $q$-degree, each of the combinations 
$x^5{\pa^3}x$, $x^4({\pa^2}x)({\pa}x)$, $x^6({\pa^2}x)\beta$, $x^9\beta^3$,
$x^6({\pa}x)({\pa}\beta)$ have $q$-degree $3$ and $t$-degree $6$.
The coefficient of $q^rt^s$ in the partition function is the number of
combinations with positive integral $q$-degree $r$ and integral
$t$-degree $s$.

Accordingly, the generating function of the zero modes 
given by powers of $x$
is a geometric series in $t$, \viz,
$1+t+t^2+t^3+\cdots=1/(1-t)$, as each monomial appears only once. 
Then, distributing $n$ derivatives $\pa$ among
$k$ number of $x$'s gives combinations with $q$-degree $n$ and $t$-degree
$k$, while distributing among $k$ number of $\beta$'s give objects with
$q$-degree $n+k$ and $t$-degree $1/t^k$. Both 
are thus counted with the number of
partitions of a positive integer $n$ in $k$ parts, $p(n,k)$. The generating
functions are, respectively, 
\begin{equation}
P_{\gamma}=\sum p(n,k) q^n t^k = \prod_{n=1}^{\infty} \frac{1}{1-q^nt}, 
\end{equation} 
and 
\begin{equation}
P_{\beta}= \sum p(n,k) q^n (q/t)^k = \prod_{n=1}^{\infty}
\frac{1}{1-q^{n+1}/t}.
\end{equation} 
The lone powers of $\beta$ without derivatives acting on
them contribute a factor $1/(1-q/t)$ to the partition function through the 
geometric series in $\beta$, similar to $x$.
Multiplying all these we obtain the partition function \eq{zc}. 
The same consideration is valid for each variable of the coordinate ring of
$\C^d$. Indeed, considering $\C[x]$ as a graded algebra of
monomials, the coordinate ring of $\C^d$ is 
$\C[x_1,x_2,\cdots ,x_d]=
\C[x_1]\otimes\C[x_2]\otimes\cdots\otimes\C[x_d]$ and the product structure
is maintained even when extended with $\beta$'s. Thus, the generating function
for a beta-gamma system on $\C^d$ is obtained as the $d$-fold product of the
generating function on $\C$, yielding \eq{zcd}. 
Let us note that it is possible to keep track of the fields by labelling 
the $t$-charges 
after the variables of the coordinate ring, if necessary, to obtain a
refinement of the partition function. 
\subsection{$\Z_N$-invariants and Molien series}
The zero mode part of the partition function  $\mathcal{Z}_{\C^d}$,
\viz, $\mathcal{Z}^{(0)}_{\C^d} =
(1-t)^{-d}$ is the Hilbert-Poincar\`e series of the graded algebra 
$\C[x_1,x_2,\cdots ,x_d]$, graded by the degree of monomials or the  
$t$-degree. Considering the action of a discrete group on the coordinate ring,
the Hilbert-Poincar\`e series of the ring of invariants is known as the
Molien series \cite{stanley,smith}. Let us consider two examples. 
 
Let $\mathcal{R}=\C[x_1,x_2]$ be the coordinate ring of $\C^2$. Let us
consider the action of the discrete group $Z_N$ on $\mathcal{R}$ as 
\begin{equation}
\label{orb:act2}
(x_1,x_2)\mapsto (\omega x_1,\omega^{-1} x_2), 
\end{equation} 
where $\omega = e^{2\pi i/N}$ denotes an $N$-th root of unity.
Then the ring of invariants is $\mathcal{R}^{\Z_N} = \C[y_1,y_2,y_3]$, with
$y_1 = x_1^N$, $y_2=x_2^N, y_3=x_1x_2$, and a syzygy $y_1y_2=y_3^N$. Since the
coordinates $y_1$ and $y_2$ are of degree $N$, $y_3$ is of degree $2$ and the
syzygy has degree of homogeneity $2N$, the Molien series is given by 
\begin{equation}
\begin{split}
\label{molien1}
\mathcal{P}_{\C^2/\Z_N} &= \frac{(1-t^{2N})}{(1-t^N)^2(1-t^2)}\\
&= \frac{(1+t^{N})}{(1-t^N)(1-t^2)}\\
\end{split}
\end{equation} 

Similarly, let us consider the three-dimensional orbifold 
$\C^3/\Z_N\times\Z_N'$, given by the action of the discrete group on the 
coordinate ring $\mathcal{R}=\C[x_1,x_2,x_3]$  of $\C^3$ as
\begin{equation}
\begin{split}
\label{orb:act3}
\Z_N &: (x_1,x_2,x_3)\mapsto (\omega x_1,\omega^{-1} x_2,x_3),\\
\Z_N' &: (x_1,x_2,x_3)\mapsto (x_1,\omega' x_2,\omega'^{-1} x_3),
\end{split}
\end{equation} 
where $\omega$ and $\omega'$ are $N$-th roots of unity. 
The ring of invariants is $\mathcal{R}^{\Z_N\times\Z_N'} =
\C[y_1,y_2,y_3,y_4]$, with $y_1=x_1^N$, $y_2=x_2^N$, $y_3=x_3^N$,
$y_4=x_1x_2x_3$, with syzygy $y_1y_2y_3=y_4^N$. Since $y_1$, $y_2$ and  $y_3$
have degree $N$, $y_4$ has degree $3$ and the syzygy is of degree $3N$, the
Molien series is given by \cite{feng}
\begin{equation}
\label{molien2}
\mathcal{P}_{\C^3/\Z_N\times\Z_N'} = \frac{(1-t^{3N})}{(1-t^N)^3(1-t^3)}.
\end{equation} 

We shall consider the partition function of 
beta-gamma system on the same orbifolds. However, imposing the syzygies 
as constraints is difficult to implement for the higher mass modes. 
The strategy of finding the partition
function for all modes is to restrict it by the orbifold action 
directly to the invariant combinations of the fields. 
Towards this goal let us note that the Molien series 
of the above two instances can be obtained as
\begin{gather}
\label{orb:quot2}
\mathcal{P}_{\C^2/\Z_N} =\frac{1}{N}\sum_{k=0}^{N-1} 
\frac{1}{(1-\omega^k t)(1-\omega^{-k}t)}\\
\label{orb:quot3}
\mathcal{P}_{\C^3/\Z_N\times\Z_N'} = 
\frac{1}{N^2}\sum_{k=0}^{N-1}\sum_{k'=0}^{N-1}
\frac{1}{(1-\omega^k t)(1-\omega^{-k} \omega'^{k'} t)(1-\omega'^{-k'} t)}.
\end{gather}
by effecting the orbifold actions \eq{orb:act2} and \eq{orb:act3}, 
respectively, on the zero mode part of the partition function \eq{zcd} and
keeping track of contribution from each coordinate.
We obtain the partition function on the orbifolds 
by projecting with the orbifold actions on the
complete partition function \eq{zcd}. 
We shall obtain formulas for more general orbifolds, by evaluating the sums.
\subsection{Sum over roots of unity as a contour integral}
\begin{figure}[h]
\begin{center}
\begin{pspicture}(-2,-2)(2,2)
\psline[linewidth=.5pt]{<->}(0,-1.7)(0,1.7)
\psline[linewidth=.5pt]{<->}(-1.7,0)(1.7,0)
\rput(1.8,.2){$\scriptstyle\mathrm{Re}(z)$}
\rput(.45,1.4){$\scriptstyle\mathrm{Im}(z)$}
\psarc[linewidth=.1pt,arrowsize=3pt 6]{c->}(1,0){.2}{0}{360}
\psarc[linewidth=.1pt,arrowsize=3pt 6]{c->}(.5,.866){.2}{0}{360}
\psarc[linewidth=.1pt,arrowsize=3pt 6]{c->}(-.5,.866){.2}{0}{360}
\psarc[linewidth=.1pt,arrowsize=3pt 6]{c->}(-1,0){.2}{0}{360}
\psarc[linewidth=.1pt,arrowsize=3pt 6]{c->}(.5,-.866){.2}{0}{360}
\psarc[linewidth=.1pt,arrowsize=3pt 6]{c->}(-.5,-.866){.2}{0}{360}
\psdots*(1,0)(.5,.866)(-.5,.866)(-1,0)(-.5,-.866)(.5,-.866)
\rput(1,.36){$\scriptstyle{C}_1$}
\rput(.9,1){$\scriptstyle{C}_2$}
\rput(-.9,1){$\scriptstyle{C}_3$}
\rput(-1,-.4){$\scriptstyle{C}_4$}
\rput(-.9,-1){$\scriptstyle{C}_5$}
\rput(1,-1){$\scriptstyle{C}_6$}
\end{pspicture}
\hskip 1cm
\begin{pspicture}(-2,-2)(2,2)
\psline[linewidth=.5pt]{<->}(0,-1.7)(0,1.7)
\psline[linewidth=.5pt]{<->}(-1.7,0)(1.7,0)
\rput(1.8,.2){$\scriptstyle\mathrm{Re}(z)$}
\rput(.45,1.4){$\scriptstyle\mathrm{Im}(z)$}
\rput(1.15,.8){$\scriptstyle\widetilde{C}$}
\psarc[linewidth=.1pt,arrowsize=3pt 6]{c->}(1,0){.2}{300}{60}
\psarc[linewidth=.1pt,arrowsize=3pt 6]{c->}(1,0){.2}{120}{240}
\psarc[linewidth=.1pt,arrowsize=3pt 6]{c->}(.5,.866){.2}{0}{120}
\psarc[linewidth=.1pt,arrowsize=3pt 6]{c->}(.5,.866){.2}{180}{300}
\psarc[linewidth=.1pt,arrowsize=3pt 6]{c->}(-.5,.866){.2}{60}{180}
\psarc[linewidth=.1pt,arrowsize=3pt 6]{c->}(-.5,.866){.2}{240}{360}
\psarc[linewidth=.1pt,arrowsize=3pt 6]{c->}(-1,0){.2}{120}{240}
\psarc[linewidth=.1pt,arrowsize=3pt 6]{c->}(-1,0){.2}{300}{60}
\psarc[linewidth=.1pt,arrowsize=3pt 6]{c->}(-.5,-.866){.2}{180}{300}
\psarc[linewidth=.1pt,arrowsize=3pt 6]{c->}(-.5,-.866){.2}{360}{120}
\psarc[linewidth=.1pt,arrowsize=3pt 6]{c->}(.5,-.866){.2}{240}{360}
\psarc[linewidth=.1pt,arrowsize=3pt 6]{c->}(.5,-.866){.2}{60}{180}
\psarc[linewidth=.1pt,arrowsize=3pt 5]{->}(0,0){1.1}{8}{52}
\psarc[linewidth=.1pt,arrowsize=3pt 5]{<-}(0,0){.9}{11}{50}
\psarc[linewidth=.1pt,arrowsize=3pt 5]{->}(0,0){1.1}{68}{112}
\psarc[linewidth=.1pt,arrowsize=3pt 5]{<-}(0,0){.9}{71}{110}
\psarc[linewidth=.1pt,arrowsize=3pt 5]{->}(0,0){1.1}{128}{172}
\psarc[linewidth=.1pt,arrowsize=3pt 5]{<-}(0,0){.9}{131}{170}
\psarc[linewidth=.1pt,arrowsize=3pt 5]{->}(0,0){1.1}{188}{232}
\psarc[linewidth=.1pt,arrowsize=3pt 5]{<-}(0,0){.9}{191}{230}
\psarc[linewidth=.1pt,arrowsize=3pt 5]{->}(0,0){1.1}{248}{292}
\psarc[linewidth=.1pt,arrowsize=3pt 5]{<-}(0,0){.9}{251}{290}
\psarc[linewidth=.1pt,arrowsize=3pt 5]{->}(0,0){1.1}{308}{352}
\psarc[linewidth=.1pt,arrowsize=3pt 5]{<-}(0,0){.9}{311}{350}
\psdots*(1,0)(.5,.866)(-.5,.866)(-1,0)(-.5,-.866)(.5,-.866)
\end{pspicture}
\hskip 1cm
\begin{pspicture}(-2,-2)(2,2)
\psline[linewidth=.5pt]{<->}(0,-1.7)(0,1.7)
\psline[linewidth=.5pt]{<->}(-1.7,0)(1.7,0)
\rput(1.8,.2){$\scriptstyle\mathrm{Re}(z)$}
\rput(.45,1.4){$\scriptstyle\mathrm{Im}(z)$}
\psarc[linewidth=.1pt,arrowsize=3pt 5]{<-}(0,0){.8}{0}{360}
\psarc[linewidth=.1pt,arrowsize=3pt 5]{->}(0,0){1.2}{0}{360}
\psdots*(1,0)(.5,.866)(-.5,.866)(-1,0)(-.5,-.866)(.5,-.866)
\rput(1,1){$\scriptstyle\mathcal{C}_2$}
\rput(.5,.4){$\scriptstyle\mathcal{C}_1$}
\end{pspicture}
\end{center}
\caption{Sequence of deformation of contour for evaluating $\mathcal{I}$
for $N=6$}
\label{fig:contour}
\end{figure}
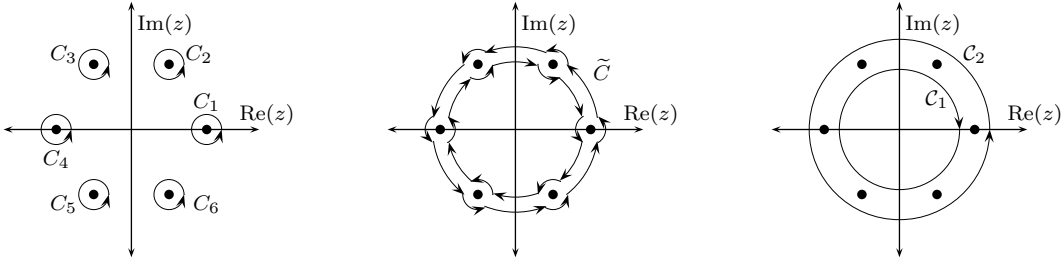

Writing the partition functions \eq{orb:quot2}, \eq{orb:quot3} as well as
their generalization to higher mass levels calls for evaluating sums over roots of 
unity. For a fixed
$N$ we need to sum over all the $N$-th roots of unity. We describe a general
procedure for evaluating such sums using elementary contour integration.
Let $g(z)$ be a function on the complex $z$-plane possessing a 
countable number of poles inside the unit circle. For a positive integer $N$
let $\omega$ denote an $N$-th root of unity. Then the sum 
\begin{equation}
\label{sum}
S=\frac{1}{N}\sum_{k=0}^{N-1} g(\omega^k)
\end{equation}
can be written as the integral
\begin{equation}
\mathcal{I}=\frac{1}{N}
\oint\limits_{\bigsqcup\limits_{k=0}^{N-1}{C}_k}\frac{g(z)}{z-\omega^k} \\ 
\end{equation} 
where ${C}_k$ are circles centered at $\omega^k$, $k=0,1,2,\cdots , N-1$,
with radii  sufficiently small so as not to
include any other pole of the integrand, as shown in the first diagram of
Figure~\ref{fig:contour}. By deforming the contour to $\tilde{\mathcal{C}}$
with two disjoint components, 
as shown in the second diagram of Figure~\ref{fig:contour}, 
we rewrite the integral as
\begin{equation} 
\mathcal{I}=\oint_{\widetilde{C}}\frac{z^{N-1}}{z^N-1}\ g(z).
\end{equation} 
The contour $\widetilde{C}$ is then deformed 
as in the third diagram of the figure
to two concentric circles $\mathcal{C}_1$ and
$\mathcal{C}_2$ with radii $1-\epsilon$ and $1+\epsilon$, respectively, with
$\epsilon\ll 1$ such that all the poles of $g(z)$ are inside the smaller circle
$\mathcal{C}_1$ and all the poles of $g(1/z)$ are outside the bigger circle
$\mathcal{C}_2$. The integral then becomes
\begin{equation} 
\mathcal{I}=\oint_{\mathcal{C}_2}\frac{z^{N-1}}{z^N-1} g(z) 
-\oint_{\mathcal{C}_1}\frac{z^{N-1}}{z^N-1} g(z). 
\end{equation} 
We then transform  $z\mapsto 1/z$ in the first integral, which is
along the outer contour $\mathcal{C}_2$, thereby exchanging the poles outside 
and inside the circle as well as changing the radius of the circle to less
than unity. The new circle may be further deformed, if necessary, to another
circle with radius $1-\epsilon'$ with $\epsilon'\ll 1$, so as to contain all 
the poles inside. Finally,  the integral becomes
\begin{equation}
\label{intI1}
\mathcal{I}= \oint_{\mathcal{C}}
\frac{dz}{z}\frac{z^N g(z)+g(1/z)}{1-z^N},  
\end{equation} 
where $\mathcal{C}$ is a circle with radius $1-\min(\epsilon,\epsilon')$, and
the poles to be taken into account are the ones inside $\mathcal{C}$ for both
$g(z)$ and $g(1/z)$, in addition to the pole at $z=0$. 

Now since  in \eq{sum}
we sum over \emph{all} the $N$-th roots of unity, we can as well 
write the sum as over the inverse of the roots of unity, that is,
\begin{equation}
S=\frac{1}{N}\sum_{k=0}^{N-1} g(\omega^{-k}).
\end{equation}
This sum can be evaluated exactly as above by considering the function 
$g(1/z)$, resulting in the expression 
\begin{equation}
\label{intI2}
\mathcal{I}= \oint_{\mathcal{C}}
\frac{dz}{z}\frac{z^N g(1/z)+g(z)}{1-z^N},  
\end{equation} 
From the two expressions \eq{intI1} and \eq{intI2} the integral can be written
in a more symmetric form as
\begin{equation}
\label{intI}
\mathcal{I}= \frac{1}{2}\oint_{\mathcal{C}}
\frac{dz}{z}\frac{1+z^N}{1-z^N}\big({g(z)+g(1/z)}\big).
\end{equation} 
We shall use this formula to evaluate the full partition function on the
orbifolds, including all the massive modes. 
\section{Partition function of beta-gamma system on $\C^2/\Z_N$}
In this section we evaluate the partition function of a beta-gamma system on
the two-dimensional orbifold $\C^2/\Z_N$ by restricting the partition
function on $\C^2$ to the invariants of the orbifold group $\Z_N$.
As discussed above, the partition function of a beta-gamma system 
on $\C^2 =\C[x_1,x_2]$ is 
\begin{equation}
\mathcal{Z}_{\C^2}  = \left(\frac{1}{(1-t)}
\prod_{n=1}^{\infty}\frac{1}{(1-q^nt)(1-q^n/t)}\right)^2,
\end{equation} 
where each factor corresponds to a coordinate in the coordinate ring. The action of the discrete group
$\Z_N$ on the coordinates is given as
\begin{equation}
(x_1, x_2) \mapsto (\omega x_1,\omega^{-a} x_2) 
\end{equation} 
with $\omega=e^{2\pi i/N}$ denoting an $N$-th root of unity and 
$1\leq a<N$ an integer. This action is lifted to the fields by embedding the $\Z_N$ into 
the $U(1)$ scaling symmetry group \eq{eq:scaling} through
identification of $\gamma$'s as $x$'s, leading to
the orbifold action
\begin{equation}
\label{a:act}
(\gamma^1, \gamma^2) \mapsto (\omega \gamma^1,\omega^{-a} \gamma^2),\quad
(\beta_1, \beta_2) \mapsto (\omega^{-1} \beta_1, \omega^a\beta_2)
\end{equation}
on the fields of the beta-gamma system.
We shall often denote the group $Z_N$ as $\frac{1}{N}(1,a)$ and
label the partition function with this alone, not mentioning $\C^2$. 
The partition function on the orbifold $\C^2/\Z_N$ is obtained as the
invariant part of $\mathcal{Z}_{\C^2}$ as
\begin{equation}
\label{pf:c2zn}
\mathcal{Z}_{\frac{1}{N}(1,a)} =\frac{1}{N}\sum_{k=0}^{N-1} 
\frac{1}{(1-\omega^k t)(1-\omega^{-ak}t)}
\prod_{n=1}^{\infty}\frac{1}{(1-\omega^k q^nt)(1-\omega^{-ak}q^nt)
(1-\omega^{-k}q^n/t)(1-\omega^{ak} q^n/t)}.
\end{equation} 
This can be thought of as a generalized Molien series extending over to the
non-zero modes. The sum is evaluated by applying \eq{intI} with 
\begin{equation}
\label{gz}
g(z) = \frac{1}{(1-zt)(1-t/z^a)}
\prod_{n=1}^{\infty}\frac{1}{(1-zq^nt)(1-z^{-a}q^nt)(1-z^{-1}q^n/t)
(1-z^aq^n/t)}.
\end{equation} 
Let us first consider the special case of unit $a$ before treating the
general case. 
\subsection{The case $a=1$}
In the special case when the discrete group $\Z_N$ acts on the
fields with $a=1$ in \eq{a:act}, the function $g(z)$ in \eq{gz} equals its
inverse, $g(1/z)=g(z)$. Using \eq{intI} this results in a simplified 
expression for the partition function, namely, 
\begin{equation}
\mathcal{Z}_{\frac{1}{N}(1,1)}= \oint_{\mathcal{C}}
\frac{dz}{z}\frac{1+z^N}{1-z^N}\ \frac{1}{(1-zt)(1-t/z)}
\prod_{n=1}^{\infty}\frac{1}{(1-zq^nt)(1-z^{-1}q^nt)(1-z^{-1}q^n/t)
(1-zq^n/t)}.
\end{equation} 
The poles to be considered in evaluating the integral 
are the ones inside the closed 
contour $\mathcal{C}$. The residue of the integrand at the simple pole 
at $z=0$ vanishes. Whether the other poles fall inside the contour depends on
the magnitude of $q$ and $t$. Let us note at this point that the expression
\eq{zc} converges  if $|t|<1$ and $|q|\ll 1$, so that $|q/t|<1$.
In this domain of the parameters 
the poles of $g(z)$ inside the contour are all simple and 
located at $z=q^n t$ and $z=q^n/t$. Evaluating residues at the poles 
leads to the infinite series for the partition function
\begin{equation}
\label{pf:11}
\mathcal{Z}_{\frac{1}{N}(1,1)}= \chi(t,q)\left(
\sum_{m=0}^{\infty} \frac{1+(q^m t)^N}{1-(q^m t)^N} \ q^{m^2}t^{2m} 
-\sum_{m=1}^{\infty} \frac{1+(q^m/t)^N}{1-(q^m/t)^N}\ q^{m^2}/t^{2m} \right),
\end{equation} 
where we defined
\begin{equation} 
\chi(t,q)= \frac{1}{1-t^2}\prod_{n=1}^{\infty} \frac{1}{(1-q^n)^2 (1-q^n
t^2)(1-q^n/t^2)}.
\end{equation} 
Rearranging  the second term  \eq{pf:11} can be written as
\begin{equation}
\label{Zc2n1}
\mathcal{Z}_{\frac{1}{N}(1,1)}= \chi(t,q)\left(
\sum_{m=0}^{\infty} \frac{1+(q^mt)^N}{1-(q^mt)^N} \ q^{m^2}t^{2m} 
+\sum_{m=1}^{\infty} \frac{1+(t/q^m)^N}{1-(t/q^m)^N}\ q^{m^2}/t^{2m}
\right).
\end{equation} 
Writing the second term of \eq{Zc2n1} as a sum over the negative integers the
partition function assumes a compact form
\begin{equation} 
\label{Zc2n2}
\mathcal{Z}_{\frac{1}{N}(1,1)}= \chi(t,q)\
\sum_{m=-\infty}^{\infty} \frac{1+(q^m t)^N}{1-(q^m t)^N}\ q^{m^2}t^{2m}.
\end{equation}  
Let us note that the change of sign of $m$ does not affect the convergence 
of the series due to the presence of $q^{m^2}$ in the numerator. 
The zero mode part, obtained as the term with $m=0$ and ignoring the
terms depending on $q$ in $\chi(q,t)$, reproduces the Molien series 
\eq{molien1}. 

In terms of theta functions with rational characteristics
\begin{equation}
\label{thetper}
\thetfn{a}{b}{\nu}{\tau} = \sum_{m=-\infty}^{\infty}
e^{i\pi\tau(m+a)^2+2i\pi (m+a)(\nu+b)} 
\end{equation} 
the partition function \eq{Zc2n2} can be expressed as
\begin{equation}
\mathcal{Z}_{\frac{1}{N}(1,1)}
= \frac{e^{i\pi(-\nu+\tau/6)}
\thetfn{0}{0}{\nu}{\tau} \thetfn{0}{0}{0}{-\tau N^2/4}
}
{i\eta(\tau/2)\thetfn{1/2}{1/2}{\nu}{\tau/2}},
\end{equation} 
if $N$ is even and as
\begin{equation} 
\mathcal{Z}_{\frac{1}{N}(1,1)}
= \frac{e^{i\pi(-\nu+\tau/6)}
\big(\thetfn{0}{0}{\nu}{\tau}\thetfn{0}{0}{0}{-\tau N^2}
+\thetfn{1/2}{0}{\nu}{\tau}\thetfn{1/2}{0}{0}{-\tau N^2}\big)
}{i\eta(\tau/2) \thetfn{1/2}{1/2}{\nu}{\tau/2}},
\end{equation} 
if $N$ is odd, where we defined $q=e^{i\pi\tau}$ and $t=e^{i\pi\nu}$.
\subsection{The case $1\leq a<N$}
As mentioned above, the general partition function in the case of $a$ being a
positive integer less than $N$ is  given by the integral 
\eq{intI} with the function $g(z)$ as given in \eq{gz}.
In this case, poles arising from both $g(z)$ and $g(1/z)$ have to be 
taken in to account, separately. Defining 
\begin{equation}
\gamma^a=t,\quad \alpha_n^a=q^nt, \quad \beta_n^a=q^n/t
\end{equation} 
first let us write $g(z)$ as 
\begin{equation} 
g(z) = \frac{1}{(1-z\gamma^a)(1-\gamma^a/z^a)}\prod_{n=1}^{\infty} 
\frac{1}{(1-z\alpha_n^a)(1-\alpha_n^a/z^a)(1-\beta_n^a/z)(1-z^a\beta_n^a)}.
\end{equation} 
Then $g(z)$ has simple poles at $z=\w^k\gamma, \w^k\alpha_n,\beta_n^a$, 
while the poles of $g(1/z)$, all simple, are located at $z=\gamma^a, 
\alpha_n^a,\w^k\beta_n$ for
every $n=1,2,\cdots ,\infty$ and $k=0,1,\cdots ,a-1$. Here
$\w$ is taken to denote an $a$-th root of unity, $\w=e^{2\pi i/a}$.
Evaluating the integral \eq{intI} using residues at these poles 
leads to the partition function 
\begin{equation} 
\label{IZ}
\mathcal{Z}_{\frac{1}{N}(1,a)}=\frac{1}{2}(Z^{(1)} + Z^{(2)}), 
\end{equation} 
where the poles of $g(z)$ contribute to the first term
\begin{equation} 
\begin{split}
\label{IZ1}
Z^{(1)} &=
\frac{1}{a}\sum_{k=0}^{a-1}\sum_{m=-\infty}^{\infty} 
\frac{1+(\w^k\alpha_m)^N}{1-(\w^k\alpha_m)^N}
\frac{1}{1-\w^k\alpha_m\gamma^{a}}
(-1)^m q^{m(m+1)/2} 
\\
&\qquad\qquad\qquad\qquad\qquad\qquad\qquad\qquad
\prod_{n=1}^{\infty} \frac{1}{(1-\w^k\alpha_m\alpha_n^a)
(1-\w^{-k}\beta_n^a/\alpha_m)(1-q^n)^2},
\end{split}
\end{equation} 
while the second term
\begin{equation} 
\begin{split}
\label{IZ2}
Z^{(2)} &= 
\frac{1}{1-t^{a+1}}
\sum_{m=-\infty}^{\infty}
\frac{1+ (q^mt)^N}{1- (q^mt)^N}
(-1)^{m(a+1)} q^{m(m+1)/2+ma(ma-1)/2} t^{ma(a+1)}
\\
&\qquad\qquad\qquad\qquad\qquad\qquad\qquad\qquad\qquad
\prod_{n=1}^{\infty}\frac{1}{(1-q^nt^{a+1})(1-q^n/t^{a+1})(1-q^n)^2}
\end{split}
\end{equation} 
contains the share of the poles of $g(1/z)$.
It appears that $Z^{(1)}$ contains fractional
powers of $q$. This is not the case due to the sum over all the $a$-th roots
of unity, which can be verified explicitly using the contour integral 
\eq{intI} once again. However, we refrain from displaying the 
expressions as they are rather cumbersome without being particularly 
illuminating otherwise.

The partition function \eq{IZ} reduces to \eq{pf:11} if we set $a=1$. 
It also matches with the direct counting 
by construction of $\Z_N$-invariant monomials to arbitrary 
given orders in $t$ and $q$.
In particular, the Molien series of some orbifolds obtained from the 
zero modes of \eq{IZ1} and \eq{IZ2} are
\begin{gather}
\mathcal{Z}_{\frac{1}{5}(1,2)}^{(0)}  = \frac{1-t+t^3}{1-t-t^5+t^6},
\\
\mathcal{Z}_{\frac{1}{7}(1,2)}^{(0)}  = \frac{1-t+t^3-t^4+t^5}{1-t-t^7+t^8},
\\
\mathcal{Z}_{\frac{1}{7}(1,3)}^{(0)}  = \frac{1-t+t^4}{1-t-t^7+t^8}\, .
\end{gather} 
\section{Partition function of beta-gamma system on $\C^3/\Z_M\times\Z_N$}
In this section we obtain the partition function of a beta-gamma system on
the threefold  $C_{MN}=\C^3/\Z_M\times\Z_N$ of $\C^3$. 
According to \eq{zcd} the partition function of a beta-gamma system 
on the affine space $\C^3 =\C[x_1,x_2,x_3]$ is, 
\begin{equation}
\label{pf:c3}
\mathcal{Z}_{\C^2}  = \left(\frac{1}{(1-t)}
\prod_{n=1}^{\infty}\frac{1}{(1-q^nt)(1-q^n/t)}\right)^3.
\end{equation} 
Let us consider the action of the  orbifold group on the coordinate ring as
\begin{gather}
\Z_M : (x_1,x_2,x_3)\mapsto (\lambda x_1,\lambda^{-1} x_2,x_3),\\
\Z_N : (x_1,x_2,x_3) \mapsto (x_1,\omega x_1,\omega^{-1} x_3),
\end{gather} 
where $\omega=e^{2\pi i/N}$ is an $N$-th root of unity and 
$\lambda=e^{2\pi i/M}$ is an $M$-th root of unity\footnote{the notation here 
is different from the previous section where $\lambda$ denoted an $a$-th 
root of unity}. As in the previous case, by embedding the discrete group
$\Z_M\times\Z_N$  into the group of scaling transformations, we have the action 
of the orbifold group on the fields, namely,
\begin{gather}
\Z_M: (\gamma^1, \gamma^2,\gamma^3) \mapsto (\lambda \gamma^1,\lambda^{-1}
\gamma^2,\gamma^3),\quad
(\beta_1, \beta_2, \beta_3) \mapsto (\lambda^{-1} \beta_1, \lambda\beta_2,
\beta_3),\\
\Z_N: (\gamma^1, \gamma^2,\gamma^3) \mapsto (\gamma^1,\omega
\gamma^2,\omega^{-1}\gamma^3),\quad
(\beta_1, \beta_2, \beta_3) \mapsto (\beta_1, \omega^{-1}\beta_2,
\omega\beta_3).
\end{gather}
Restriction of the partition function \eq{pf:c3} by 
the orbifold group to the invariant part is
\begin{equation}
\begin{split}
\mathcal{Z}_{C_{MN}} = \frac{1}{MN}\sum_{k=0}^{M-1}&\sum_{r=0}^{N-1}
\frac{1}{(1-\lambda^kt)(1-\lambda^{-k}\omega^rt)(1-\omega^{-r}t)} \\
&\prod_{n=1}^{\infty} \frac{1}{(1-\lambda^kq^nt)(1-\lambda^{-k}\omega^rq^nt)
(1-\omega^{-r}q^nt)} \\
&\prod_{n=1}^{\infty} \frac{1}{(1-\lambda^{-k}q^n/t)
(1-\lambda^{k}\omega^{-r}q^n/t)(1-\omega^{r}q^n/t)}.
\end{split}
\end{equation}
The sums over the two sets of roots of unity can be performed in two steps. 
Let us define $g$ as a function of two complex variables $\zeta$ and $z$, 
namely,
\begin{equation}
\begin{split}
g(\zeta,z) = 
\frac{1}{(1-t\zeta )(1-tz/\zeta)(1-t/z)} 
&\prod_{n=1}^{\infty} \frac{1}{(1-q^nt\zeta)(1-q^ntz/\zeta)
(1-q^nt/z)} \\
&\prod_{n=1}^{\infty} \frac{1}{(1-q^n/t\zeta)
(1-q^n\zeta/tz)(1-zq^n/t)}.
\end{split}
\end{equation}
First we evaluate the sum over the $N$-th roots of unity, using \eq{intI} 
with $g(\zeta,z)$ treated as a function of $z$. This is done by working out the
contour integral \eq{intI} over $z$ by evaluating the residues of 
$g(\zeta,z)$ as well as
$g(\zeta,1/z)$ as in the previous section to arrive at the partition function
written as a sum over powers of $\lambda$, namely,
\begin{equation} 
\label{ZM}
\mathcal{Z}_{C_{MN}} = \frac{1}{2M}\sum_{k=0}^{M-1} 
\chi_{\zeta}(t,q)
\sum^{\infty}_{m=-\infty} 
\left( 
\frac{1+(q^mt)^N}{1-(q^mt)^N}
+\frac{1+({q^mt}/{\zeta})^N}{1-({q^mt}/{\zeta})^N}
\right) 
\left.\frac{q^{m^2}t^{2m}}{\zeta^m}\right|_{\zeta=\lambda^k},
\end{equation} 
where we defined
\begin{equation}
\chi_{\zeta}(t,q)= \frac{1}{(1-\zeta t)(1-\frac{t^2}{\zeta})}\prod_{n=1}^{\infty} 
\frac{1}{
(1-q^n)^2 (1-\zeta q^nt) (1-\frac{q^nt^2}{\zeta}) (1-\frac{q^n}{\zeta t}) 
(1-\frac{q^n\zeta}{t^2})
}.
\end{equation} 
In order to evaluate the sum over the $M$-th roots of unity 
we  employ the contour integration 
formula \eq{intI} once again with the summand 
in \eq{ZM} for $g$ now treated as a function of $\zeta$. 
Now the integrand has simple poles  arising from 
$\chi_{\zeta}(t,q)$, as before. However, the second term of \eq{ZM} possesses 
additional simple poles, located 
at $\zeta = \omega^k q^mt$, for $k=0,1,\cdots , N-1$ and every integer $m$. 
The contribution from the former set of poles to the partition function adds
up to
\begin{equation}
\label{Z1:ser}
\mathcal{Z}_1 = \frac{1}{4}\chi_0 \!\!\sum_{m,n=-\infty}^{\infty}
\left(
\frac{1+(q^mt)^N}{1-(q^mt)^N}
+\frac{1+(q^{m+n}t^2)^N}{1-(q^{m+n} t^2)^N}
\right) 
\frac{1+(q^nt)^M}{1-(q^nt)^M}\ 
q^{m^2+mn+n^2}t^{3(m+n)},
\end{equation} 
where we have defined 
\begin{equation}
\chi_0  =  \frac{1}{1-t^3}\prod_{n=1}^{\infty} 
\frac{1}{(1-q^n)^4(1-q^nt^3)(1-q^n/t^3)} 
\end{equation} 
The latter set of poles lead to another sum over the $N$-th roots of 
unity, as
\begin{equation}
\label{Nav}
\mathcal{Z}_2 = \frac{1}{2N} 
\sum_{k=0}^{N-1}
\chi_{z}(t,q)
\sum_{m=-\infty}^{\infty}
\frac{1+(q^m t/z)^M}{1-(q^m t/z)^M}
\left.\frac{q^{m^2}t^{2m}}{z^m}\right|_{z=\omega^k}.
\end{equation} 
In terms of these two parts partition function now assumes the form 
\begin{equation} 
\label{Z12}
\mathcal{Z}_{C_{M,N}} =\mathcal{Z}_1+\mathcal{Z}_2.
\end{equation} 
Evidently, the use of the contour integration method again 
to evaluate $\mathcal{Z}_2$ in \eq{Nav} is beleaguered by the 
existence of poles of the integrand at $z=\lambda^k q^mt$, 
which brings back a sum over the $M$-th roots of unity. 
In order to obtain a series not involving the roots of unity
we use the identity \cite{kac}
\begin{equation} 
\chi_{\zeta}(t,q) = \chi_0 \left(
\sum_{m=0}^{\infty} \sum_{n=0}^{\infty}
q^{mn} \zeta^{n-m} t^{n+2m}
-\sum_{m=1}^{\infty} \sum_{n=1}^{\infty}
q^{mn} \zeta^{n-m} t^{-(m+2n)}
\right)
\end{equation} 
in \eq{Nav} to expand $\chi_z(t,q)$ in powers of $z$. 
The sum over the $N$-th roots of unity in \eq{Nav} to restrict to the 
invariant part can now
be performed by keeping only the terms in which powers of $z$ vanish 
modulo $N$. This leads to a series expansion of $\mathcal{Z}_2$ as
\begin{equation}
\begin{split}
\label{Z2:ser}
\mathcal{Z}_{2} &=
\chi_0\sum_{m=0}^{\infty} \sum_{n=0}^{\infty} \sum_{p=0}^{\infty} 
\sum_{s=-\infty}^{[(m+n+Mp)/N]}
q^{m^2+n^2+mn+Mp(m+n)-Nms} t^{3(m+n)+2Mp-Ns}\\      
&+
\chi_0\sum_{m=1}^{\infty} \sum_{n=1}^{\infty} \sum_{p=0}^{\infty} 
\sum_{s=-\infty}^{[(m+n+Mp-1)/N]}
q^{m^2+n^2+mn+Mp(m+n)-Nms} t^{-(3(m+n)+2Mp-Ns)}\\   
&-
\chi_0\sum_{m=1}^{\infty} \sum_{n=0}^{\infty} \sum_{p=0}^{\infty} 
\sum_{s=-\infty}^{[(m+n+Mp-1)/N]}
q^{m^2+n^2+mn+Mp(m+n)-Nms} t^{-(3m+Mp-2Ns)}\\       
&-
\chi_0\sum_{m=0}^{\infty} \sum_{n=1}^{\infty} \sum_{p=0}^{\infty} 
\sum_{s=-\infty}^{[(m+n+Mp)/N]}
q^{m^2+n^2+mn+Mp(m+n)-Nms} t^{3m+Mp-2Ns}\\          
&-
\frac{1}{2}\chi_0\sum_{m=0}^{\infty} \sum_{n=0}^{\infty} 
\sum_{s=-\infty}^{[(m+n)/N]}
q^{m^2+n^2+mn-Nms} t^{3(m+n)-Ns}\\                  
&-
\frac{1}{2}\chi_0\sum_{m=1}^{\infty} \sum_{n=1}^{\infty} 
\sum_{s=-\infty}^{[(m+n-1)/N]}
q^{m^2+n^2+mn-Nms} t^{-(3(m+n)-Ns)}\\               
&+
\frac{1}{2}\chi_0\sum_{m=1}^{\infty} \sum_{n=0}^{\infty} 
\sum_{s=-\infty}^{[(m+n-1)/N]}
q^{m^2+n^2+mn-Nms} t^{-(3m-2Ns)}\\                  
&+
\frac{1}{2}\chi_0\sum_{m=0}^{\infty} \sum_{n=1}^{\infty} 
\sum_{s=-\infty}^{[(m+n)/N]}
q^{m^2+n^2+mn-Nms} t^{3m-2Ns},                  
\end{split}
\end{equation} 
where $[r]$ denotes the largest integer less than a rational number $r$.
Using this expression in \eq{Z12} along with \eq{Z1:ser} gives a series 
for the partition function in terms of $q$ and $t$.
\subsection{Molien series from zero modes}
Molien series of various orbifolds can be evaluated from 
\eq{Z12} by collecting the zero mode parts from \eq{Z1:ser} and \eq{Z2:ser}.
Here is a list of  examples 
\begin{gather}
Z^{(0)}_{C_{2,2}}  = \frac{1-t+t^2}{(1-t)^3 (1+t)^2},
\quad
Z^{(0)}_{C_{2,3}} = \frac{1-t+2t^3-t^5+t^6}{(1-t)^3 (1+t)^2
   (1+t^2+t^4)},
\quad
Z^{(0)}_{C_{3,3}} = \frac{1-t^9}{(1-t^3)^4},
\nonumber\\
Z^{(0)}_{C_{5,5}} = 
\frac{t^8-t^7+t^5-t^4+t^3-t+1}{(1-t)^3
   (t^4+t^3+t^2+t+1)^2},
\\
Z^{(0)}_{C_{3,5}} = 
-\frac{\left(t^4-t^2+1\right) \left((t-1) t
   \left(t^3+t+1\right)+1\right) \left(t
   \left(t^4-t^3+t-1\right)+1\right)}{t \left(t
   \left((t-1)^2 t^{13}-1\right)+2\right)-1}.
\nonumber
\end{gather} 
These expressions match with the direct computation of invariant monomials.  
\section{Conclusion}
We have obtained explicit expressions for the partition function of curved
beta-gamma system on orbifolds of affine spaces in two and three dimensions,
namely, $\C^2/\Z_N$ and $\C^3/\Z_M\times\Z_N$. 
Since an affine space can be covered with a single chart, the fields $\gamma$
of a free beta-gamma system 
can be identified with the indeterminates of the coordinate ring of the affine
space. This in turn allows lifting the geometric action of the orbifold 
group to the fields. Its action on the conjugates is then determined by 
requiring the invariance of the action under scaling. The partition function 
on the orbifold is then obtained by restricting the partition function on the 
affine spaces to the part invariant under the orbifold group. 
The partition function so obtained is a generalized
Molien series in the sense that its zero mode part furnishes the Molien
series of the orbifolds. While the Molien series may be evaluated by
other means, e.g., using the syzygies describing the orbifolds as
algebraic varieties in the coordinate rings of the affine spaces, imposing
the constraint on the full partition function turns out to be 
computationally cumbersome at the least. Restriction to the invariant sector 
on the other hand, entails summing over roots of unity.
For simple cases the summation may be performed order 
by order  in $q$. This however becomes extremely involved even for $N$ bigger 
than $2$ for $\C^2/\Z_N$, for instance, rendering the task of
obtaining formulas at higher orders in $q$  difficult. 
We evaluated the sum by expressing it as a contour integral in a single
complex variable. While this elementary treatment suffices for the surfaces,
the partition function on the 
threefolds involves sums over two sets of roots of unity. In that case 
a combination of a contour integration and an
infinite series identity is employed to obtain a series for the partition
function. We note that the contour integration employed above can straightforwardly be 
generalized to higher dimensions.
A series for the partition function obtained after taking care of summation 
over the roots of unity is effective for its explicit evaluation. We hope the
formulas presented to be of use in obtaining physical quantities for
beta-gamma systems. In particular, evaluation of various moments of the
partition function, which yield different physical quantities, 
will be facilitated.


\end{document}